\documentclass{aastex} 
\usepackage{amssymb,amsmath,graphicx,psfrag,mathtools}
\usepackage{url}
\usepackage{revsymb}
\begin{document}
\title{Log-Normal Widths of Waiting Time Distributions: A Fundamental Parameter}
\author{J. I. Katz}
\affil{Department of Physics and McDonnell Center for the Space Sciences\\
Washington University, St. Louis, Mo. 63130}
\email{katz@wuphys.wustl.edu}
\begin{abstract}
	Many astronomical phenomena, including Fast Radio Bursts and Soft
	Gamma Repeaters, consist of brief, separated, seemingly aperiodic
	events.  The intervals between these events vary randomly, but there
	are epochs of greater activity, with shorter mean intervals, and of
	lesser activity, with longer mean intervals.  {This variability can
	be quantified by} a single dimensionless parameter, the width of a
	log-normal fit to the distribution of waiting times between events.
	If the distribution of event strengths is a power law, {as is
	often the case,} this parameter is independent of the detection
	threshold, and is a robust measure of the {intrinsic variability
	of the waiting times} {and of the underlying dynamics}.
\end{abstract}
\section{Introduction}
Many episodic natural phenomena originate in complex and imperfectly
understood physical processes {\it e.g.,} {repeating} Fast Radio
Bursts (FRB) \citep{Z23} and Soft Gamma Repeaters (SGR) \citep{H11}.  In
some, such as FRB, the responsible physical processes and their environment
are not known.  In others we may know which physical processes are
responsible (magnetic reconnection in SGR), but lack sufficient
understanding to calculate them quantitatively.

Burst activity may be described by the distribution of intervals (waiting
times) between detected bursts.  This reflects two distinct properties of
the underlying {process}: the correlation (or lack thereof) between
consecutive bursts (short-term memory) and slower variations in the mean
level of activity (long-term memory).  {When the level of activity
varies, the width of the distribution increases; SGR 1806$-$20 is an example
(Table \ref{data}).}  The distributions of waiting times in
repeating FRB \citep{K19,A22,H22,L21,N22,Zha22} and in SGR burst storms
\citep{H94,Y20} have been well fit by log-normal functions.  This paper
describes the application of log-normal fits to waiting time distributions,
and suggests their dimensionless standard deviation $\sigma$ as a robust and
fundamental quantitative metric.  Alternative statistical methods have
recently been used to describe FRB waiting times by \citet{DWSX,WSX23}.
Their conclusions about memory in burst recurrence intervals (``short
waiting times tend to be followed by short ones, and long by long'') are
qualitatively consistent with the picture of varying levels of activity
presented here.
\section{Bursting Phenomena}
Bursting phenomena may be described by their distribution of strengths
(energy, flux, fluence, electromagnetic field) and by their 
temporal distribution.  Distributions of strengths are power laws
\citep{Y20,Zha22} {when there is no characteristic value of strength
(earthquake magnitude, flare or burst energy, {\it etc.\/}) or scale over a
wide range} \citep{K41a,K41b,K86}.  Any deviation from a power law would
define a characteristic {strength}, the value of {strength} where
its distribution deviates from a straight line on a log-log plot,
{contradicting the assumption that there is no characteristic strength}.
In some FRB a break {(a change in the exponent of the fitted energy
distribution $f(E)$)} is observed \citep{Zha22}, defining a characteristic
value {of the burst energy $E$ or fluence where the exponent changes},
and some SGR have shown extreme outliers, also not described by a power law
distribution \citep{K21}; power laws are widespread but not universal.  {As
a result, many but not all log-normal widths are independent of detection
sensitivity.}


Most episodically outbursting phenomena display periods of greater and
lesser activity with shorter and longer mean waiting times, respectively.
{If the activity level is uniform the process is, by definition,
Poissonian and the width of the waiting time distribution $\sigma = 0.723$,
while if there are periods of greater and lesser activity there are excesses
of short and long waiting times, leading to a broader distribution of
waiting times.} The width $\sigma$ {of this distribution} thus measures
the {\it variability\/} of the activity.  In one limit outbursts are
regularly periodic and $\sigma = 0$, while in the other limit brief periods
of frequent activity are separated by long periods of quietude and $\sigma
\gg 1$; shot noise, with random events occurring at a constant mean rate, is
intermediate.
\section{Log-Normal Fits}
Log-normal fitting functions are widely used because with only three
parameters, a midpoint, a maximum and a width, they provide good fits to a
broad range of single-peaked distributions, wide as well as narrow
\citep{JKB94,LSA01}.  These fits may be entirely phenomenological, without
basis in a causal model.  Most uses of log-normal fits describe the
distribution of some parameter of the individual events, but here we are
concerned with the waiting times between them.

The logarithm of the product of a series of multiplications of a random
variable is the sum of the logarithms of the variables, and executes a
random walk.  If there are many factors in the product the central limit
theorem applies to the sum of the logarithms, and the result approaches a
log-normal distribution \citep{S57,MS82}.  The rate of convergence depends
on the distributions of these logarithms, and for pathological distributions
the sum may not converge at all.

A log-normal distribution of a variable is therefore a natural consequence
if it is the result of a series of independent stochastic multiplicative
steps.  The observed $\sigma$ of a waiting time distribution would be
$\propto N^{-1/2}$, where $N$ is the number of independent steps, in series,
weighted by their frequency: if there is a necessary step that occurs at a
low rate, the observed waiting time distribution would reflect the
distribution of that infrequent step, while a step that occurs at a high
rate would have little effect on $\sigma$.  A log-normal distribution will
be a good fit if there are several required steps, each with a comparable
rate, making the central limit theorem applicable to the sum of their
logarithms.

Power-law distributions are not fit by log-normal functions because they do
not have a peak.  All power-law distributions of data must have at least one
cutoff, either a threshold for detection or an intrinsic characteristic
scale, in order that the total number of events and the total energy (or an
analogous quantity) be finite.  The distribution peaks at the cutoff
{because it marks the transition between an increasing function that would
diverge if not cut off and a decreasing function}.
Abruptly cut off power laws may be fit by log-normal functions, although not
closely, {and the present method may be applied}.

{A log-normally distributed probability density function of $N$
measurements of the variable $x$, {normalized to unity,} is
\begin{equation}
	\label{x}
	f(x) = {\exp{[-(\ln{x}-\ln{x_0})^2/2\sigma^2]} \over
	x\sqrt{2\pi}\sigma},
\end{equation}
where $\ln{x_0}$ is the mean logarithm and $\sigma$ the standard deviation
of the logarithms \citep{EHP,JKB94}.  Applying this to} a distribution of
waiting times $\Delta t$ between events, the distribution of their
logarithms, {normalized to $N$ events,} is
\begin{equation}
	\label{Deltat}
	f(\ln{\Delta t}) = A \exp{[-(\ln{\Delta t}-\ln{\Delta t_0})^2/
	2 \sigma^2]},
\end{equation}
where $\ln{\Delta t_0}$ is the {peak of the distribution of logarithms} and
$\sigma$ is its dimensionless standard deviation.  If there are $N$ values
of $\Delta t$ in the data the normalization factor $A = N/(\sqrt{2\pi}
\sigma)$.  {We are interested in $\sigma$ as a parameter that describes
the underlying physical processes.}

For data with the empirical distribution $f(\ln{\Delta t})$,
$\ln{\Delta t_0}$ is taken to be the mean
\begin{equation}
	\label{Dt0}
	\ln{\Delta t}_0 = {\int_{-\infty}^\infty\!f(\ln{\Delta t})
	\ln{\Delta t}\, d\ln{\Delta t} \over \int_{-\infty}^\infty\!
	f(\ln{\Delta t})\,d\ln{\Delta t}}.
\end{equation}
The standard deviation
\begin{equation}
	\label{sigma}
	\sigma = \sqrt{{1 \over \pi}{\int_{-\infty}^\infty\!f(\ln{\Delta t})
	(\ln{\Delta t} - \ln{\Delta t}_0)^2\,d\ln{\Delta t} \over
	\int_{-\infty}^\infty\!f(\ln{\Delta t})\,d\ln{\Delta t}}}.
\end{equation}
\section{Simple Models}
\subsection{Exact Periodicity}
One limiting case of a waiting time distribution is that of periodic pulses,
{all of which are bright enough to be detected}; all waiting times are
then the same.  This is {often} a excellent approximation for radio
pulsars, whose period derivatives $10^{-21} \lesssim {\dot P} \lesssim
10^{-9}$.  Then {the fractional variation in inter-pulse intervals over
a span $T$ of observations} (generally, very intermittent rather than
continuous) is $T{\dot P}/P$ and the fitted $\sigma \sim T {\dot P}/P$.
For observed pulsars $\sigma$ is in the range $10^{-14}$--0.03 \citep{PSR},
with the largest values for young, rapidly slowing, pulsars (like the Crab)
observed for decades, and the smallest values for a recycled (low-field)
pulsar briefly observed.  For a steadily slowing but nearly strictly
periodic phenomenon like pulsar emission, $\sigma$ is not very meaningful
because of its dependence on $T$.
\subsection{Shot Noise}
{The definition of} shot noise (Poissonian statistics) with mean rate
$\nu$ {is that the probability per unit time of an event after a waiting
time $\Delta t$ from the immediately preceding event decays exponentially at
the rate $\nu$:}
\begin{equation}
	f(\Delta t) = \nu\Delta t \exp{(-\nu\Delta t)} = \exp{\left[
		\ln{\nu\Delta t} - \exp{(\ln{\nu\Delta t})}\right]}.
\end{equation}
Performing the integrals {from 0 to $\infty$} in Eq.~\ref{Dt0}
\begin{equation}
	\ln{(\nu\Delta t)_0} = -0.577,
\end{equation}
and Eq.~\ref{sigma}
\begin{equation}
	\sigma = 0.723.
\end{equation}
\subsection{Two Widely Separated $\Delta t$}
\label{two}
Another limiting case is that of only two possible values of $\Delta t$,
$\Delta t_1$ and $\Delta t_2$:
\begin{equation}
	f(\ln{\Delta t}) = a \delta(\ln{\Delta t} - \ln{\Delta t}_1) +
	(1-a) \delta(\ln{\Delta t} - \ln{\Delta t}_2).
\end{equation}
Then 
\begin{equation}
	\ln{\Delta t}_0 = a \ln{\Delta t}_1 + (1-a) \ln{\Delta t}_2
\end{equation}
and
\begin{equation}
	\sigma = \sqrt{{1 \over \pi}
	\left[a\left(\ln{({\Delta t}_1/{\Delta t}_0)}\right)^2
	+ (1-a)\left(\ln{({\Delta t}_2/{\Delta t}_0)}\right)^2\right]}.
\end{equation}
This is the limit of a distribution with two narrow but separated peaks.  

Double-peaked distributions are frequently found for the waiting times
between fast radio bursts \citep{K19,L21,Z21,A22,H22,N22,Zha22,J23}.  The
peaks at shorter waiting times may be attributable to substructure
within individual bursts rather than to the intervals between distinct
bursts, and are not considered further.
\section{Log-Normal Fit Examples}
{Log-normal widths have been fitted to the waiting time distributions of
many astronomical object with episodic activity.  This section collects the
best-characterized examples.  The object must have discrete, cleanly
separated, events such as FRB or SGR outbursts, and enough events must have
been observed ($\gtrsim 100$) to determine $\sigma$ with smalll uncertainty.
Solar flares display memory on many time scales \citep{AJ21} but are beyond
the scope of this study.}

Table~\ref{data} shows the values of $\sigma$ of waiting time distributions
of shot noise and of several astronomical datasets.  These include very
active FRB, burst storms from SGR 1806$-$20 and SGR 1935$+$2154 (associated
with a Galactic FRB) and microglitches of the Vela pulsar. 

\begin{table}
	\centering
	\begin{tabular}{|lrcl|}
		\hline
		Process or Object&$N$ & $\sigma$ & \\
		\hline
		Periodic & $\infty$ & 0 & \\
		Shot noise & $\infty$ & 0.723 & \\
		FRB 20121102 & 1652 & $1.27 \pm 0.02$ & \citet{L21}\\
		FRB 20121102 & 133 & $1.12 \pm 0.07$ & \citet{A22}\\
		FRB 20121102 & 478 & $1.22 \pm 0.04$ & \citet{H22}\\
		FRB 20201124A & 996 & $1.04 \pm 0.03$ & \citet{N22}\\
		FRB 20201124A & $> 800$ & $1.33 \pm 0.03$ & \citet{Zha22}\\
		SGR 1806$-$20 & 111 & $3.46 \pm 0.23$ & \citet{H94}\\
		SGR 1806$-$20 & 262 & $3.6\phantom{0} \pm 0.13$ & \citet{G00}\\
		SGR 1935$+$2154 & 217 & $1.08 \pm 0.05$ & \citet{Y20}\\
		Four flare stars & 88--205 & $0.683 \pm 0.015$ & \citet{W23}\\
		Vela PSR $\mu$-glitches & 57 & $0.79 \pm 0.07$ & \citet{H94}\\
		Vela PSR $\delta{\dot\nu}$ & 57 & $0.33 \pm 0.03$ & \citet{H94}\\
		\hline
	\end{tabular}
	\caption{\label{data} Values of $\sigma$ {obtained for
	log-normal fits to} several datasets.  $N$
	{is the number of events in each dataset.}  The uncertainties of
	$\sigma$ are their standard errors based on the number of data
	(where FRB intervals are evidently double-peaked the peak at longer
	intervals is used).  \citet{H94} gave values of $\sigma$ but the
	other $\sigma$ (and that for Vela PSR $\delta {\dot\nu}$) were
	measured from the published figures.  The FRB (see Table 6 of
	\citet{HH22} for a review), SGR and flare star data refer to the
	intervals (waiting times) between bursts or flares.  {For FRB
	the data are averages over several observing epochs by the same
	telescope.}  The Vela PSR intervals are the waiting times between
	its microglitches and $\delta {\dot \nu}$ the changes in its
	spindown rate following these glitches.  The flare star data are
	averages over targets 176.01, 218.01, 1224.01 and 1450.01 observed
	{continuously} by TESS \citep{W23}; the uncertainty cited is the
	standard error of the mean $\sigma$.  \citet{N22} and \citet{Zha22}
	are based on the same observing campaign.}
\end{table}

{The values of $\sigma$ in the Table indicate the width of a log normal
function fitted to the empirical distribution of waiting times for these
datasets, and its values for a periodic signal (like a pulsar) and for shot
noise.  Smaller $\sigma$ indicate a narrower distribution of waiting times
(the distribution is a $\delta$-function for a strictly periodic signal) and
larger values indicate a wider distribution, with intervals distributed over
a wide range.  The values of $\sigma$ in the table depend on the exclusion
of anomalously short intervals that are attributed to substructure within a
single event that may be erroneously classified as distinct outbursts.  The
criteria for exclusion are necessarily subjective, but plausible variations
change $\sigma$ by only a few hundredths.  For SGR 1806-20 {$\sigma$ was
calculated by the cited references.} The intervals range over several orders
of magnitude---it has intense burst storms separated by years of quiescence,
{explaining the large value of $\sigma$}.  Other $\sigma$ were
calculated from the cited published data.}
\section{Discussion}
This paper compares the widths of log-normal fits to interval (lag) times
of several different kinds of bursting phenomena, observed over several
decades by different instruments.  The data and analyses are necessarily
heterogeneous.  When the original authors provided uncertainty estimates,
those are used.  When they are not available, uncertainties have been
estimated from the published figures, as indicated in the Table caption.
The data and their analyses are described in the papers cited here.

Inspection of these figures (the reader is referred to the cited original
papers) indicates that the log-normal functions fit the data well, but it
is not possible to say with confidence whether deviations from the
log-normal are entirely explained by the random statistics of comparatively
small samples.  Even if this is not the case, the log-normals explain most
of the distribution of waiting times, and are a valid characterization, the
the dominant terms in a moment expansion about the mean.  It is not claimed
that the log-normal functions describe the data exactly (aside from the
statistics of small samples) but that, considered as a leading term in an
expansion, they contain fundamental information about the underlying
physical processes.

The virtue of $\sigma$ as a measure of the {statistics of} a bursting
source is its independence of the observing sensitivity if the distribution
of event strengths is a power law.  This is simply demonstrated: If $\sigma$
depended on a detection threshold $\cal T$, then the function
$\sigma({\cal T})$ would define the characteristic signal strength equal to
$\cal T$ where $\sigma$ is some specific value, such as 0.723 (its value for
shot noise).  That would be inconsistent with a power law distribution of
signal strength, a straight line on a log-log plot of number of events {\it
vs.\/} signal strength, with no characteristic value \citep{K41a,K41b,K86}.
Most of the phenomena under consideration (astronomical fast radio bursts,
soft gamma repeaters and pulsar glitches) do not have obvious characteristic
strengths, at least over a range several orders of magnitude wide, but there
have been exceptions \citep{Zha22}.

Very small ($\ll 1$) values of $\sigma$ of a waiting time distribution
indicate periodicity.  Values $<0.723$, the value for shot noise, indicate a
memory effect, like that of a noisy relaxation oscillator, that does not
produce periodicity but instead has a characteristic repetition time scale,
with shorter or longer waiting times less likely than for shot noise.  This
is often described as quasi-periodicity, and is shown in Table~\ref{data}
for the post-microglitch changes of the spin-down rate of the Vela pulsar.
These $\delta{\dot \nu}$ have a preferred scale, although the timing of the
microglitches is consistent with shot noise.  {True periodicity is
described by $\sigma \to 0$.  Although this limit cannot be achieved in a
finite dataset, phenomena that are called periodic, such as binary orbits or
pulsar rotation, have nonzero but very small $\sigma \ll 1$.}

Values of $\sigma > 0.723$ indicate varying rates of activity (shot noise
has the most random possible statistics if the mean or statistically
expected rate of activity is unchanging).  Most of the data series shown in
Table~\ref{data} have $\sigma$ larger than 0.723, indicating changing levels
of activity and long term memory in the mechanism.  In contrast, the smaller
value of $\sigma$ for the Vela $\delta{\dot\nu}$ indicates a characteristic
scale of the phenomenon (because these data are not time intervals, it
doesn't indicate periodicity of quasi-periodicity).

Much larger values of $\sigma$, as found for SGR 1806$-$20, may indicate
periods of much greater (``burst storms'') and of lesser activity; long
intervals are likely to be followed and preceded by long intervals, and
short intervals by short intervals.  The fact that SGR and FRB show periods
of greater and lesser activity has long been known; \citet{Zho22} show an
extreme example for FRB 20201124A, that resumed activity after an extended
quiet period \citep{Niu}.  A different toy model is discussed in
Sec.~\ref{two}, in which intervals may be either long or short, but there is
no memory bunching long intervals together, nor short intervals together.
The fact that $\sigma$ of SGR1935$+$2154, the only SGR associated with a
FRB, is much less than that of SGR 1806$-$20 indicates that these are
fundamentally different objects; unfortunately, insufficient data are
available from other SGR to explore this further.

Table~\ref{data} shows that for the repeating FRB for which sufficient
waiting time data are available, $\sigma$ has approximately the same value,
indicating common underlying {processes}.  Few other repeating FRB have
sufficient data to determine $\sigma$ (unfortunately, the Galactic FRB
200428, remarkably associated with a SGR, does not).  In contrast, the two
SGR waiting time distributions have widths whose difference far exceeds
their uncertainties; perhaps SGR 1935$+$2154, the source of FRB 200428, is
fundamentally different from other SGR, even the most intense of which has
not been associated with a FRB \citep{TKP}.

{The fact that different observations of the same source during periods
with different mean levels of activity (this applies both to FRB and to the
one SGR for which two different datasets exist) are described by very
similar values of $\sigma$ indicates that this is an intrinsic property of
these objects.  The fact that two different FRB have similar $\sigma$
suggests that repeating FRB are qualitatively similar, despite their 
very great differences in repetition rate and other properties; data from
more repeating FRB will be required to test this hypothesis.}  {This
conclusion does not require that the values of $\sigma$ agree within their
small nominal one-standard deviation uncertainties, but only that they
differ by $\ll 1$.  For example, the fact that values of $\sigma$ for FRB
are about 2.5 less than those of SGR 1806$-$20 is sufficient to demonstrate
that they are not produced by objects like SGR 1806$-$20, and the fact that
they are about 0.5 greater that that of shot noise demonstrates that the
FRB have memory.}

The mean value of $\sigma$ for the flare stars is close to that of shot
noise, as is that for Vela micro-glitches.  Unlike the other processes
considered, these appear to be uncorrelated random events.  This is
consistent with these events occuring as the result of localized processes
in distinct uncoupled small regions, of the stellar photosphere for flare
stars and of the neutron star interior for micro-glitches.  The flare star
data comprise 104, 205, 88 and 91 flares over periods of 1035 d, 1853 d,
1848 d and 1266 d, respectively, for TESS targets 176.01, 218.01, 1224.01
and 1450.01.  There is no statistical evidence for cycles of activity
analogous to the Solar cycle, but this may be attributable to the limited
temporal extent of the data.

The universality classes of critical phenomena \citep{PV02} suggest a path
to qualitative categorization.  Phenomena of different microphysical origin
are sorted into universality classes on the basis of their critical
exponents.  Detailed knowledge of their microphysics, such as intermolecular
potentials in liquid-gas critical points or the Hamiltonian of a
ferromagnet, is not required to identify these classes and to establish
fundamentally common dynamics among phenomena involving different physical
processes.  This paper suggests that the width of a log-normal distribution
is analogous to the critical exponents that characterize universality
classes in renormalization group theory.
\section*{Acknowledgment}
I thank N. Whitsett for permission to use unpublished data and D. Eardley,
J. Goodman, R. Grober, M. Maharbiz, W. Press, A. Rollett and J. Tonry for
useful discussions.
\section*{Data Availability}
This theoretical study did not generate any new data.

\label{lastpage} 
\end{document}